\documentclass[%
 aps,
 jmp,%
 amsmath,amssymb,
 twocolumn,
 showpacs
]{revtex4-1}

\usepackage{graphicx}
\usepackage{dcolumn}
\usepackage{bm}
\usepackage{color}

\begin{document}

\preprint{AIP/123-QED}

\title[Sample title]{
Quantum Monte Carlo study of the formation of molecular polarizations and \\
the antiferroelectric ordering in squaric acid crystals 
}

\author{Hiroaki Ishizuka}
\author{Yukitoshi Motome}%
\affiliation{ 
Department of Applied Physics, University of Tokyo, 
Tokyo 113-8656, Japan
}%

\author{Nobuo Furukawa$^{1,2}$}
\author{Sei Suzuki$^1$}
\affiliation{%
$^1$Department of Physics and Mathematics, Aoyama Gakuin University, 
Kanagawa 252-5258, Japan
\\
$^2$Multiferroics Project, ERATO, Japan Science and Technology Agency (JST), 
c/o Department of Applied Physics, University of Tokyo, Tokyo 113-8656, Japan 
}%

\date{\today}

\begin{abstract}
Effects of geometrical frustration and quantum fluctuation are theoretically investigated for the 
proton ordering in a quasi-two-dimensional hydrogen-bonded system, squaric acid crystal. 
We elucidate the phase diagram for an effective model, the transverse-field Ising model on a frustrated checkerboard
lattice, by using quantum Monte Carlo simulation.
A crossover to liquidlike paraelectric state with well-developed molecular polarizations is
identified, distinguishably from long-range ordering.
Emergence of long-range order from the liquidlike state exhibits peculiar aspects 
originating from the lifting of quasi-macroscopic degeneracy, 
such as colossal enhancement of the transition temperature and a vanishingly small anomaly in the specific heat. 
\end{abstract}

\maketitle

\section{ \label{sec:intro}
Introduction
}
Proton ordering in hydrogen-bonded systems has long been one of the central topics in condensed matter physics. 
Each proton is in a double-minimum potential on the hydrogen bond, 
and spatial correlations among the proton configurations strongly affect macroscopic properties of hydrogen-bonded crystals. 
A famous example is the ``ice-rule" configuration of protons  and its relation to the residual entropy in water ice~\cite{Bernal1933,Pauling1933}. 
Another example is the ferroelectricity due to the proton ordering in KH$_2$PO$_4$~\cite{Slater1941,Blinc1960,deGennes1963}. 
Electronic polarization emerging from proton ordering has been extensively studied with emphasis
on both the fundamental physics and the application to electronic devices~\cite{Lines1977,Horiuchi2008}.

\begin{figure}
  \begin{center}
  \includegraphics[width=3.4in]{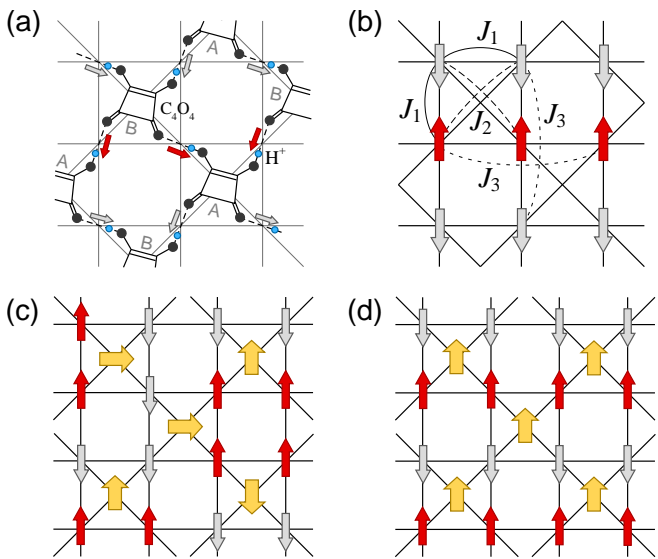}
  \end{center}
  \caption{
  (Color online) 
  (a) Schematic picture of a layer of H$_2$SQ molecules.
  The dashed lines represent hydrogen bonds connecting C$_4$O$_4$ units, and 
  small circles on the bonds show protons. 
  (b) Pseudospin representation of the proton displacement in (a). 
  $J_1$, $J_2$, and $J_3$ denote the interactions in Eq.~(\ref{eq:hamiltonian}).
  (c) An example of intermediate liquidlike states in which 
  every molecule bears a polarization but the system is globally disordered. 
  (d) A ferroelectrically-ordered state. 
  In (c) and (d), the bold arrows in the center of plaquettes represent
  the molecular polarizations. See the text for details. 
  }
  \label{fig:sqacid}
\end{figure}

Here, we focus on one of such hydrogen-bonded materials, squaric acid crystal H$_2$C$_4$O$_4$ (H$_2$SQ).
H$_2$SQ is a quasi-two-dimensional (2D) molecular solid.  In each 2D layer, squaric acid molecules form a network of hydrogen bonds, 
as shown in Fig.~\ref{fig:sqacid}(a)~\cite{Semmingsen1973}. 
A particular configuration of protons induces a polarization in each molecule, 
and an ordering of the polarizations can lead to ferroelectricity. 
In fact, H$_2$SQ exhibits antiferroelectricity below $T_{\rm c}=375$K, which is driven by the 2D ferroelectric ordering 
with interlayer antiferroelectric coupling~\cite{Samuelsen1977}. 

H$_2$SQ has two striking aspects. 
One is the local constraint on the proton positions similar to the ice rule in water ice. 
Each molecule has four hydrogen bonds; two out of four protons come close to the C$_4$O$_4$ unit and the other two are far. 
The local constraint alone is not sufficient to determine a unique ground state 
and brings about a macroscopic degeneracy, as seen in water ice; H$_2$SQ has geometrical frustration in nature. 

The other aspect is the effect of quantum tunneling of protons. 
In general, the external pressure increases the tunneling rate of protons between two potential minima, which reduces local polarizations, 
and consequently, suppresses the ferroelectricity.  
Indeed, in H$_2$SQ, $T_{\rm c}$ is suppressed with increasing pressure. 
A peculiar intermediate state, however, appears before the polarization is lost in each molecule: 
the macroscopic polarization vanishes while the polarization in each molecule is retained~\cite{Moritomo1991_3}.
Consequently, a quantum paraelectric state is realized in the low-temperature limit. 

There have been many theoretical studies for the antiferroelectric transition in H$_2$SQ. 
The local constraint and associated frustration were considered on the basis of vertex models or 
frustrated pseudospin models~\cite{Deininghaus1981,Stilck1981,Zinenko1976,Matsushita1980}. 
The coupling between pseudospins and phonons was also studied~\cite{Chaudhuri1990,Wesselinowa1995,Dalal1998}. 
Most of the studies, however, were limited at the mean-field level,
and the effect of geometrical frustration has not been fully clarified yet. 
In particular, quantum fluctuation under the geometrical frustration,
which is presumably important for understanding the quantum paraelectricity under pressure,
has not been seriously considered so far. 

In the present study, we investigate an effective model for H$_2$SQ, a
2D checkerboard-lattice Ising model with transverse field
which corresponds to the application of external pressure. 
With a sophisticated quantum Monte Carlo (QMC) method, we map out the numerically-exact phase diagram.
We identify a liquidlike state intervening between the ferroelectric phase and the paraelectric phase. 
In the intermediate state, molecular polarizations are well retained by ice-rule type local correlations,
but they are globally disordered. 
The peculiar nature of transition from the intermediate state to ferroelectric phase is discussed.

The organization of this paper is as follows. In Sec.~\ref{sec:model_and_method}, we introduce models and methods.
After introducing the pseudo-spin model in Sec.~\ref{sec:model}, we explain the numerical method and the definition of 
the observables in Sec.~\ref{sec:method} and Sec.~\ref{sec:order}, respectively.
The results of calculation are presented in Sec.~\ref{sec:results}.
The temperature dependence of observables is given in Sec.~\ref{sec:dependency}
and the phase diagram in Sec.~\ref{sec:diagram}.
Discussions on our results with the previous studies are elaborated in Sec.~\ref{sec:discussions}.
Section~\ref{sec:summary} is devoted to summary.

\section{ \label{sec:model_and_method}
Model and method
}

\subsection{ \label{sec:model}
Pseudo-spin model
}

We here consider a pseudospin model for H$_2$SQ 
following the previous studies~\cite{Zinenko1976,Matsushita1980}. 
In the pseudospin model, proton displacements 
in a plane of the bipartite square lattice of H$_2$SQ molecules [Fig.~\ref{fig:sqacid}(a)] 
are represented by the $z$-component of pseudospins 
as $\sigma_i^z = \pm 1$ [Fig.~\ref{fig:sqacid}(b)]~\cite{Blinc1960,deGennes1963}. 
Here the appropriate signs are assigned so that protons belonging to A(B) sublattice H$_2$SQ molecules correspond to the up(down) spins. 
The local constraint similar to the ice rule in water ice~\cite{Bernal1933,Pauling1933} (two out of four protons are close and the other two are far)
is taken into account by the antiferromagnetic interactions between nearest neighbors, $J_1$, and crisscrossing next-nearest neighbors, $J_2$, 
on the checkerboard lattice~\cite{Zinenko1976,Matsushita1980}. 
When $J_1 = J_2$, the model is a 2D variant of the spin-ice model~\cite{Harris1997,Ramirez1999}, 
in which sixfold degeneracy in each plaquette results in a macroscopic number of energetically-degenerate ground states $\sim 1.5^{N/2}$.
In the present case, we take $J_2 > J_1$ since two closer protons favor an edge of the C$_4$O$_4$ square,
not a diagonal~\cite{Zinenko1976,Matsushita1980}. 
$J_2$ larger than $J_1$ partially lifts the degeneracy, but the ground-state degeneracy still remains: 
all configurations with different stacking of antiferromagnetic 1D diagonal chains, 
exemplified in Figs.~\ref{fig:sqacid}(c) and \ref{fig:sqacid}(d), give the same lowest energy. 
The degeneracy is reduced but still quasi-macroscopic, i.e., $4^{\sqrt{N}}$. 
Hence, the present $J_1$-$J_2$ model does not show any long-range order down to zero $T$~\cite{Deininghaus1981,Stilck1981}, 
although a finite-$T$ transition was discussed in the previous mean-field studies~\cite{Zinenko1976,Matsushita1980}. 

To stabilize a ferroelectric ordering [a stripe ordering in terms of pseudospins, shown in Fig.~\ref{fig:sqacid}(d)], 
a further degeneracy-lifting perturbation must be included. 
In the present study, we consider an inter-molecular coupling originating from the distortion by forming C=C double bonds. 
A C=C double bond induces a trapezoid-type distortion of C$_4$ square, and favors 
a ferro-type alignment of the molecules since a short C=C bond tends to elongate the neighboring parallel C-C bonds in the adjacent molecules. 
The inter-molecular correlation is incorporated by a third-neighbor ferromagnetic interaction $J_3$ [Fig.~\ref{fig:sqacid}(b)]. 
$J_3$ lifts the remaining degeneracy and selects the ferroelectrically-ordered state. 

By summarizing the above argument, our model is given by the following Hamiltonian
\begin{eqnarray}
{\cal H}= J_1 \! \sum_{\langle i, j \rangle} \! \sigma_i^z \sigma_j^z
 + J_2 \! \sum_{\left[ i,j \right]}  \! \sigma_i^z \sigma_j^z
 - J_3 \! \! \sum_{ \left\{ i,j \right\} }  \! \sigma_i^z \sigma_j^z
 + \Gamma \sum_{i} \! \sigma_i^x, 
 \label{eq:hamiltonian}
\end{eqnarray}
where, $\sigma_i^\alpha$ is the $\alpha$ component of the Pauli matrix, 
representing the pseudospin operator at site $i$; $J_1, J_2, J_3 > 0$, and 
the first, second, and third sums are taken between the nearest, second (crisscrossing), 
third neighbors on the checkerboard lattice, as shown in Fig.~\ref{fig:sqacid}(b). 
Here, the last term with the transverse field $\Gamma$ is introduced 
to represent the quantum tunneling of protons. 
We take $J_1=1$ as the energy unit, and focus on the case with $J_2=2$. 
The results are qualitatively the same for $J_2 > J_1$. 
We analyze the 2D model by changing $J_3$, $\Gamma$, and $T$ 
to clarify the nature of inplane ferroelectricity in H$_2$SQ; 
the effect of the interplane coupling will be 
mentioned later. 

\subsection{ \label{sec:method}
Quantum Monte Carlo method
}

To investigate the thermodynamics of the model given by Eq.~(\ref{eq:hamiltonian}), 
we employ a recently-developed continuous-time QMC method with a cluster update in the imaginary-time direction~\cite{Nakamura2008}. 
In order to overcome the slow relaxation in the present frustrated system, 
we use the replica exchange method~\cite{Fukushima1996} and the loop-flip update algorithm~\cite{Rahman1972}.
In the replica exchange, the replicas are chosen along the constant $\Gamma/T$ lines 
since the Boltzmann weight strongly depends on the numbers of domain walls 
along the imaginary-time direction which are proportional to $\Gamma/T$. 
For the loop-flip algorithm, we applied two variants of the original algorithm.
One is the original algorithm which forms a loop by connecting up spins and down spins alternatively,
with treating the  $J_1$ and $J_2$ bonds equivalently.
The other one is the diagonal flipping process; a diagonal chain of spins connected by
$J_2$ is flipped at once.
The system sizes are taken from $L = 24$ to $36$ ($N=4L^2$) under periodic boundary conditions. 
Typically, MC measurements are performed for 100000 samplings after 30000 initial thermalizations. 
Results are divided into six bins to estimate statistical errors by the variance among the bins.

\subsection{ \label{sec:order}
Physical quantities
}

To distinguish the long-range order and ``ice-rule" type local correlations, 
we calculate the macroscopic polarization $P$ and local correlation parameter
$\rho$. $P$ is calculated as 
\begin{eqnarray}
P = [S(0,\pi)^2 + S(\pi,0)^2]^{1/2}
\end{eqnarray}
via the spin structure factor 
\begin{eqnarray}
S({\bf k}) = \frac1N \sum_{i,j} \sigma_i^z \sigma_j^z \exp(-i{\bf k} \cdot {\bf r}_{ij}), 
\end{eqnarray}
which detects the stripe-type ordering in Fig.~\ref{fig:sqacid}(d). 
The critical temperature is determined by the Binder analysis~\cite{Binder1981} using the Binder parameter for
$P$,
\begin{eqnarray}
g_P = \frac12 \left( 3 - \frac{\langle P^4 \rangle}{\langle P^2 \rangle^2} \right). \label{eq:binder}
\end{eqnarray}
On the other hand, $\rho$ detects the fourfold-degenerate stable configuration in each plaquette by 
\begin{eqnarray}
\rho = \frac2N \sum_{p} f(p),
\end{eqnarray}
where the sum $p$ runs over all the crisscrossing plaquettes, and 
$f(p)$ is a function giving 1 for the fourfold stable states and otherwise $-1/3$ in $p$th plaquette: $\rho \to 0$ 
when the spins are completely disordered, and $\rho \to 1$ when all the plaquettes are in the fourfold stable states. 
Note that $\rho$ is not an order parameter but characterizes a crossover
associated with the formation of molecular polarizations as demonstrated below.
We also measure the corresponding susceptibilities, $\chi_P$ and $\chi_\rho$, 
from the fluctuations of $P$ and $\rho$ as
\begin{eqnarray}
\chi_P    &=& \frac{N}{T}(\langle P^2\rangle - \langle P \rangle^2), \\
\chi_\rho &=& \frac{N}{T}(\langle \rho^2\rangle - \langle \rho \rangle^2),
\end{eqnarray}
respectively. The specific heat is calculated by 
\begin{eqnarray}
C = \frac{1}{NT^2} ( \langle {\cal H}^2\rangle - \langle {\cal H} \rangle^2 ).
\end{eqnarray}

\section{\label{sec:results}
Results
}

\subsection{\label{sec:dependency}
Locally-correlated liquidlike state
}

\begin{figure}
  \begin{center}
  \includegraphics[width=3.0in]{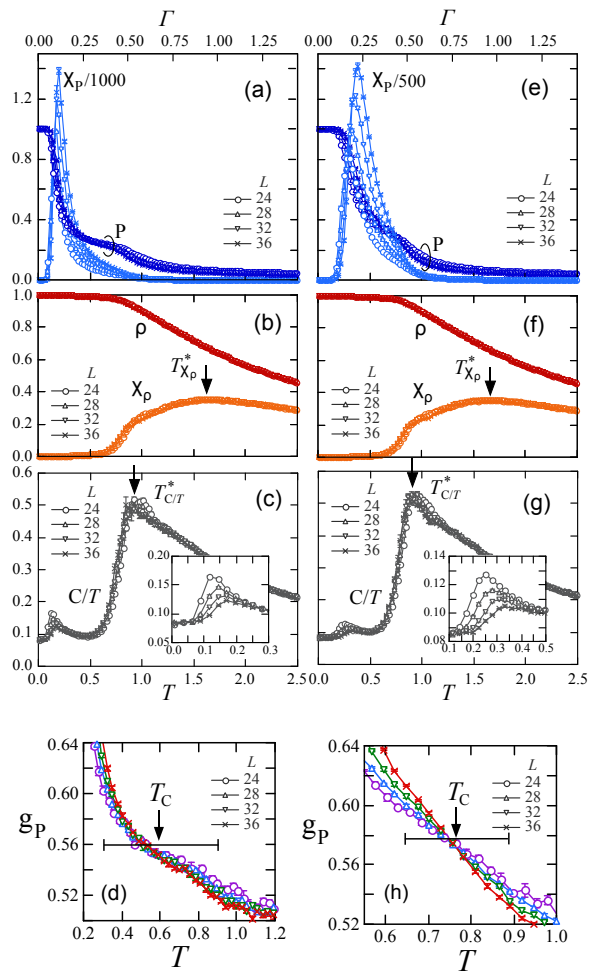}
  \end{center}
   \caption{
   (Color online) QMC results of (a)(e) $P$ and $\chi_P$ ($\chi_P$ is divided by 1000 and 500, respectively), 
   (b)(f) $\rho$ and $\chi_\rho$, and (c)(g) $C$ for the system sizes $N = 4L^2$ ranging from $L=24$ to $L=36$.
   The data are at (a)-(c) $J_3=0.002$ and (e)-(g) $J_3=0.004$. 
   (d) and (h) shows the Binder parameter in vicinity of the $T_c$ for $J_3=0.002$ and $J_3=0.004$, respectively.
   Insets of (c)(g) show the low-$T$ behaviors of $C/T$. 
   All the results are calculated at $J_2=2$ along the axis with $\Gamma/T=\tan(\pi/6)$.
   See the text for details. 
   }
   \label{fig:size}
\end{figure}

Figure~\ref{fig:size} shows QMC results along $\Gamma/T=\tan(\pi/6)$ axis at $J_3=0.002$ and $0.004$.
At the lowest $T$, the system shows a ferroelectric ordering with fully-saturated polarization $P$,
as shown in Figs.~\ref{fig:size}(a) and \ref{fig:size}(e). 
With increasing $T$ and $\Gamma$, $P$ steeply decreases and the corresponding susceptibility $\chi_P$ shows 
a sharp peak which grows as $N$ increases, indicating a phase transition into a paraelectric state. 
The critical temperature $T_{\rm c}$ is estimated from the crossing point of the Binder parameter of $P$ given by Eq.~(\ref{eq:binder}), 
as shown in Figs.~\ref{fig:size}(d) and \ref{fig:size}(h): 
$T_{\rm c} = 0.60(30)$ for $J_3=0.002$ and $T_{\rm c} = 0.76(12)$ for $J_3=0.004$.
On the other hand, the local correlation parameter $\rho$ 
remains to be large even above $T_{\rm c}$ and gradually decreases for $T \gtrsim 1.0$ [Figs.~\ref{fig:size}(b) and \ref{fig:size}(f)]. 
Correspondingly, the specific heat $C$ divided by $T$ has a peak, as shown in Figs.~\ref{fig:size}(c) and
\ref{fig:size}(g). The susceptibility for $\rho$, $\chi_\rho$, shows a broad peak at a higher $T$. These indicate
that the system does not directly enter into a completely disordered state at $T_{\rm c}$ but exhibits an
intermediate state in which molecular polarizations are retained; the system shows a crossover to a
completely-disordered paraelectric state characterized by the peak of $C/T$ or $\chi_\rho$ at $T_{C/T}^*$ or
$T_{\chi_\rho}^*$. 
The intermediate state is a liquidlike paraelectric state originating from the ice-rule type local correlations. 

In the ferroelectric phase transition and crossover to liquidlike phase, most of the entropy is released at the
crossover by forming the ice-rule type manifold. As shown in Figs.~\ref{fig:size}(c) and \ref{fig:size}(g), $C/T$
sharply decreases with the saturation of $\rho$, and the peak associated with the phase transition at a lower $T$
is very small. In fact, as shown in the insets, the intensity of the small peak decreases as $N$ increases, while
the peak position approaches $T_{\rm c}$. This suggests that the entropy associated with the phase transition
becomes vanishingly small in the thermodynamic limit. The peculiar behavior is understood by considering the
quasi-macroscopic degeneracy in the ice-rule type manifold where the remaining entropy is in the order of
$\sqrt{N}$ not $N$~\cite{Stilck1981}.

\begin{figure}
  \begin{center}
  \includegraphics[width=3.0in]{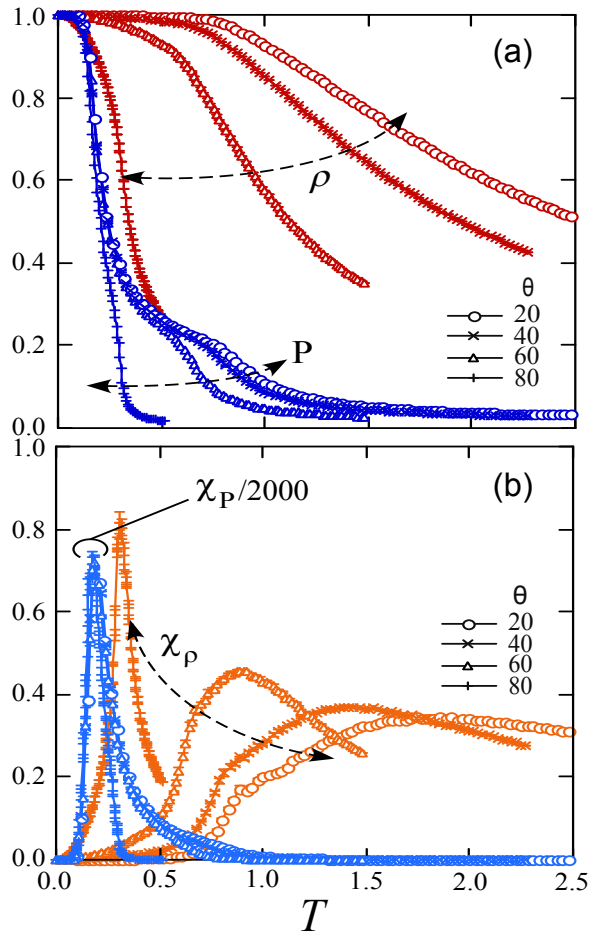}
  \end{center}
   \caption{
   (Color online)
     $T$ dependences of (a) $P$ and $\rho$, and (b) $\chi_P/2000$ and $\chi_\rho$, 
     measured along various $\Gamma/T$. 
     $\Gamma/T$ is parametrized by $\theta$ (degree) with $\Gamma/T = \tan\theta$. 
    All the results are calculated at $J_2=2$ and $J_3=0.002$ for $N=4\times 36^2$ sites.
   }
   \label{fig:angle}
\end{figure}

The intermediate liquidlike state is widely observed while changing $\Gamma/T$. 
Figure~\ref{fig:angle} shows 
$P$, $\rho$, and their susceptibilities along the various $\Gamma/T$ axes. 
As indicated by a decrease of $\rho$ and a broad peak of $\chi_\rho$, 
the crossover temperature $T^*$ largely decreases with increasing $\Gamma$, 
while the data of $P$ and $\chi_P$ show that $T_{\rm c}$ does not decreases so rapidly. 
There, however, is always a window of the intermediate state 
with well-developed local correlations and suppressed global order.

\subsection{\label{sec:diagram}
Phase diagrams
}

\begin{figure}
  \begin{center}
  \includegraphics[width=2.94in]{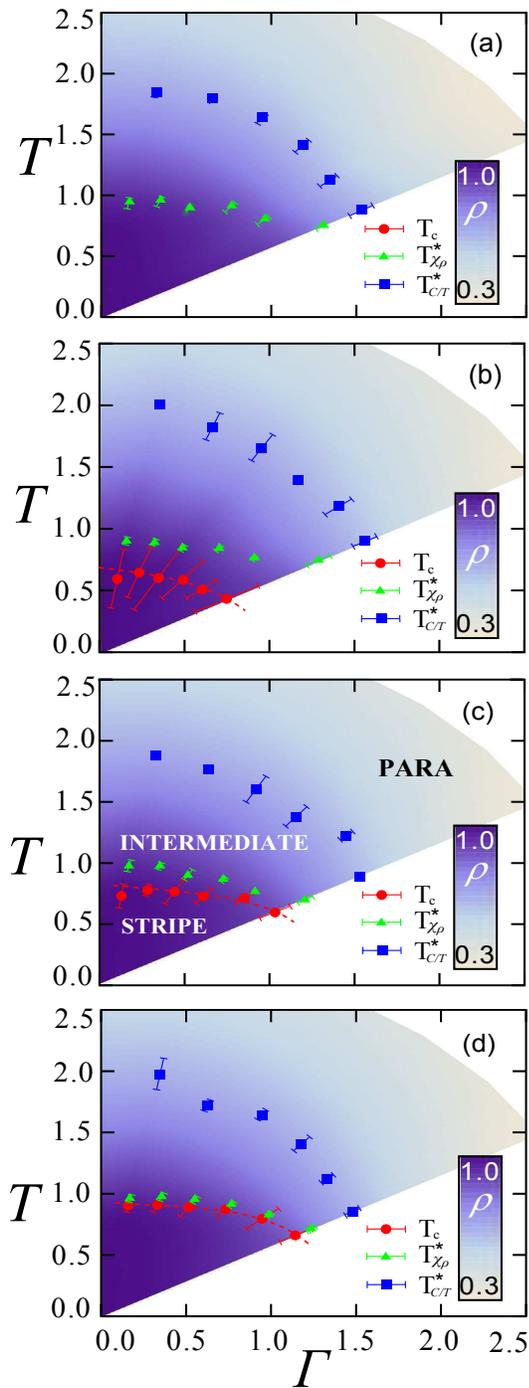}
  \end{center}
  \caption{
    (Color online) Phase diagrams obtained for varying $J_3$.
    Each diagram for (a)-(d) corresponds to $J_3=0, 0.002, 0.004, 0.008$, respectively.
    The solid circles shows $T_c$ obtained by Binder analysis and triangular (square) points are $T^\ast$
    evaluated from $C/T$ ($\chi_\rho$).
    The dotted line shows phase boundary of ordered state and paraelectric state, and
    the intermediate state is assigned as the region between $T_{\chi_\rho}^\ast$ and $T_c$.
    The gradation shows intensity of $\rho$.
  }
  \label{fig:diagram}
\end{figure}

The phase diagrams for $\Gamma$, $T$, and $J_3$ are summarized in Fig.~\ref{fig:diagram}. 
$(T_{\rm c}, \Gamma_{\rm c})$ are estimated by the Binder analysis of $P$,
and $(T^*, \Gamma^*)$ are identified by a peak of $\chi_\rho$ or $C/T$. 
In the case of $J_3=0$, the system exhibits only the crossover into the liquidlike state with quasi-macroscopic degeneracy
[the region below $T_{\chi_\rho}^\ast$; see Figs. 2(b) and (f)], and remains paraelectric down to the lowest $T$ calculated. 
One might expect a quantum order by disorder as predicted for the case of 
$J_1=J_2$~\cite{Moessner2001}, but it will be limited to very low $T$ region,
if any, and is out of the scope of the present study.
With $J_3$ switched on, the degeneracy is lifted and the ferroelecrically-ordered phase emerges inside the liquidlike state. 
The ordered state rapidly extends with increasing $J_3$, but a sequential
change of the three different regimes is clearly observed in the region where
$J_3$ is sufficiently small. With further increasing $J_3$, $T_{\text{c}}$
exceeds $T^\ast$ and the intermediate liquidlike state is taken over by the long-range ordered state.

A remarkable point of the phase diagram is the rapid growth of $T_{\rm c}$ with $J_3$,
which is more than 100 times larger than $J_3$.
This colossal enhancement of $T_{\rm c}$ is also ascribed to the peculiar nature of the
intermediate state with quasi-macroscopic degeneracy; 
the spatial correlations are strongly enhanced along the diagonal chains.
This strong correlation reinforces the effect of $J_3$ as $J_2$ and $J_3$ are not frustrated,
and the system becomes extremely sensitive to the degeneracy-lifting perturbation $J_3$. 
Another point is the fate of the intermediate state at low temperatures.
Since the QMC simulation becomes harder for larger $\Gamma/T$,
it is difficult to conclude whether the intermediate state remains at low temperatures.
Nonetheless, we expect a finite window down to low $T$ for small $J_3$ from
the systematic change of the phase diagram shown in Fig. 4.

\section{\label{sec:discussions}
Discussions
}

Comparing with experiments, we successfully identify the liquidlike state with well-developed molecular
polarizations, which might account for the peculiar intermediate state in experiments~\cite{Moritomo1991_3}. 
Our results suggest that the intriguing physics related with the ice-rule type degeneracy is involved within the 2D layers of H$_2$SQ crystal. 
We note, however, that the qualitative shape of the phase diagram appears to be different; 
in experiments, both $T_{\rm c}$ and $T^*$ decrease almost linearly in applied pressure and their difference is
almost independent of pressure. This can be ascribed to the parametrization of the realistic situation; i.e., how
the model parameters change under pressure. The first-principle calculations may help further quantitative
studies~\cite{Rovira2001}. Another possible origin of the discrepancy is an ambiguity in assigning $T^*$;
in general, crossover boundary depends on how to define or detect it. With regard to the nature of the phase
transition, our results indicate that it is second order and the associated anomaly of the specific heat is
vanishingly small. These apparent contradictions with experiments~\cite{Barth1979,Kuhn1979,Mehring1981} might
be reconciled by considering the interlayer coupling or more complicated couplings to lattice distortions,
which are neglected in our model~\cite{Mehring1981-2,Wang1989,Wesselinowa1995}.

Comparing with the previous theoretical researches, our results are obtained by seriously including both
geometrical frustration and quantum fluctuation, which were not fully taken into account in the previous
studies. For instance, in the absence of $J_3$, our result shows no phase transition as expected in the frustrated
situation, in contrast to the previous research in which a finite-temperature phase transition was predicted
because of the mean-field type treatment. Furthermore, our result clearly indicated existence of the liquidlike
state as well as  several significant consequences of the local correlation on the thermodynamical properties.
In particular, the absence of anomaly in the specific heat at $T_c$ and the strong enhancement of $T_c$ by $J_3$
are revealed for the first time by our calculations.
On the other hand, as mentioned above, our result also showed continuous transition as in many previous
theoretical studies using pseudospin or an equivalent approach. This suggests that an extension of the model
is necessary to account for the first-order transition in experiments.

\section{\label{sec:summary}
Summary
}

To summarize, we have investigated the geometrically-frustrated transverse-field Ising model as an effective
model for squaric acid crystals. By unbiased QMC simulations, we have identified an intermediate liquidlike
state between the ferroelectrically-ordered state and the completely-disordered paraelectric state, in which
molecular polarizations are well preserved but they are globally disordered due to the frustration.
Furthermore, we found out that the emergence of locally correlated state significantly affects the thermodynamic
behavior of this system. In particular, we unveiled the vanishingly-small anomaly in the specific heat at the
transition point and the colossal enhancement of $T_c$ by the degeneracy-lifting perturbation $J_3$. Emergence
of such state and the remarkable effects have never been reported in the previous theories.
The liquidlike state accounts for a peculiar intermediate paraelectric state observed under external pressure
in the squaric acid crystal.

Although our results qualitatively reproduce the peculiar phase diagram of the squaric acid under
pressure, the quantitative changes of the critical temperature and the crossover temperature are different
from those in experiments. Further quantitative researches are necessary. For example, experimentally,
detailed analysis of the structure under pressure will be quite important for more quantitative comparison
between experiment and theory. Moreover, the first-principle calculation will help to identify the model
parameters and their changes under pressure more precisely. On the other hand, our result exhibits second
order phase transition, which is in contrast to the first order transition observed experimentally. This might
require further extension of the model, for example, by including the interlayer coupling and complicated
couplings to lattice distortions. Such extension is left for the future work.

\section*{Acknowledgement}

The authors thank Y. Tokura for fruitful discussions.
H.I. and Y.M. thank T. Misawa, Y. Motoyama, H. Shinaoka, and M. Udagawa for helpful comments.
This research was supported by KAKENHI (No. 19052008, No. 20740225, and No. 22540372), and 
Global COE Program ``the Physical Sciences Frontier."

\nocite{*}
\bibliography{aipsamp}

\end{document}